\documentclass[twocolumn,showpacs,amsmath,amssymb,prl,superscriptaddress]{revtex4}
\usepackage{graphicx}% Include figure files
\usepackage{bm}% bold math
\usepackage{psfig}

\begin{document}

\title{Strain-induced Fermi contour anisotropy of GaAs 2D holes}

\author{J.~Shabani}
\affiliation{Department of Electrical Engineering,
Princeton University, Princeton, NJ 08544, USA}
\author{M.~Shayegan}
\affiliation{Department of Electrical Engineering,
Princeton University, Princeton, NJ 08544, USA}
\author{R.~Winkler}
\affiliation{Department of Physics, Northern Illinois University,
DeKalb, IL 60115, USA }

\date{\today}

\begin{abstract}
We report measurements of magneto-resistance commensurability peaks,
induced by a square array of anti-dots, in GaAs (311)A
two-dimensional holes as a function of applied in-plane strain. The
data directly probe the shapes of the Fermi contours of the two spin
subbands that are split thanks to the spin-orbit interaction and
strain. The experimental results are in quantitative agreement with
the predictions of accurate energy band calculations, and reveal
that the majority spin-subband has a severely distorted Fermi
contour whose anisotropy can be tuned with strain.

\end{abstract}


\maketitle

%%%%%%%%%%%%%%%%%%%%%%%%%%%%%%%%%%%%%%%%%%%%%%%%%%%%%%%%%%%%%%%%%%

Manipulation of the spin-orbit coupling and the resulting spin-splitting
in two-dimensional (2D) carrier systems is of considerable current
interest \cite{Winkler03}. Such systems are the basis for novel spintronic
devices \cite{DattaAPL90,Reviews} and also allow studies of fundamental phenomena such as various phases that
the 2D carriers acquire in mesoscopic structures
\cite{MorpurgoPRL98,YauPRL02,KogaPRB04}, and the spin Hall effect
\cite{EngelsCM06}. An important problem is the ballistic transport
of the carriers in these systems and, in particular, the ability
to resolve and manipulate such transport for the two
spin-subbands. Measurements of the commensurability oscillations
\cite{LuPRL98} and, more recently, magnetic focusing
\cite{RokhinsonPRL04}, have indeed revealed spin-resolved
ballistic transport in GaAs 2D hole systems (2DHSs). Here we
report spin-resolved ballistic transport measurements in GaAs
2DHSs as a function of applied in-plane strain. Together with the results of our energy band calculations,
the data reveal a strong distortion of the spin-subband Fermi contours with
strain, especially for the majority subband. Such distortion could
find use in ballistic spintronic devices that rely on the spatial
separation of carriers with different spin \cite{RokhinsonPRL04}.

Figure 1 captures the essence of our study. In Fig.\ 1(a) we show
the calculated Fermi contours of the heavy and light heavy-hole
(HHh and HHl) spin-subbands for a GaAs (311)A 2DHS at different
values of strain ($\epsilon$) applied along the [$01\bar{1}$]
direction. There is a severe distortion of the contours,
especially for the HHh band, as a function of strain. Such
distortions have been theoretically reported previously but
experimental evidence has been only through piezo-resistance
measurements which qualitatively agree with the calculations
\cite{KolokolovPRB99,HabibAPL07}. In our study, we perforated the
2DHS with anti-dot (AD) lattices, and measured its
magneto-resistance (MR) in an L-shaped Hall bar mesa. In the AD
lattice, the carriers move along real-space cyclotron trajectories
with orbit shapes that are similar to the shape of the Fermi
contour in reciprocal space but are rotated by 90$^\circ$. When
these orbits become ``commensurate'' with the AD lattice period, the
MR exhibits a peak \cite{Weiss91, Lorke91, Fleisch92, TsukagoshiPRB95,
LuPRB96, OkiPRB07, ZitzlspergerEPL03}.
The primary (smallest) such orbits are shown in Fig.\ 1(b) for the
HHh and HHl holes for the current parallel to the [$01\bar{1}$]
direction.  The MR peaks measured along [$01\bar{1}$] and
[$\bar{2}33$] should then reveal direct information regarding the
size and shape of the HHh and HHl contours. Here we provide such
MR data and compare them with the peak positions expected from the
calculated Fermi contours.
\begin{figure}
\centering
\includegraphics[scale=1]{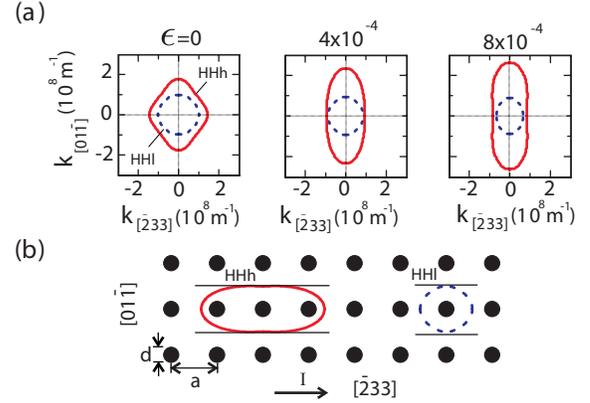}
\caption{\label{fig:fig1}(Color online) (a) Calculated Fermi
contours for the HHh and HHl bands of GaAs (311)A 2D holes at
$p=2.45 \times 10^{11}$ cm$^{-2}$ at the indicated strain
($\epsilon$) values. (b) The primary commensurate orbits for the
HHh and HHl bands for $\epsilon=5 \times 10^{-4}$ for current
along $[\bar{2}33]$ (see text).}
\end{figure}

Our sample was grown on a GaAs (311)A substrate by molecular beam
epitaxy and contains a modulation-doped 2DHS confined to a
GaAs/AlGaAs heterojunction. The Al$_{0.35}$Ga$_{0.65}$As/GaAs
interface is 100~nm below the surface and is separated from a
17~nm-thick Si-doped Al$_{0.35}$Ga$_{0.65}$As layer by a 30~nm
Al$_{0.35}$Ga$_{0.65}$As spacer layer. The 2DHS has a typical low
temperature mobility of $1 \times 10^5$~cm$^2$/Vs along
$[01\bar{1}]$ and $4.3 \times 10^5$~cm$^2$/Vs along $[\bar{2}33]$ at
a 2D hole density, $p$, of $2.9 \times 10^{11}$~cm$^{-2}$. We
fabricated an L-shaped Hall bar, with its arms along $[01\bar{1}]$
and $[\bar{2}33]$, via optical photo-lithography. We then deposited
a layer of PMMA and patterned the square AD arrays using electron
beam lithography. The AD pattern was etched to a depth of $\simeq
50$~nm; this depth is sufficiently large to strip the dopant layer
and thus deplete the holes in the AD regions, but small enough to
avoid introducing non-uniform strain in the AD lattice. In our
sample there are two AD lattice regions, with periods $a=600$ and
750 nm, patterned on the $[01\bar{1}]$ arm of the Hall bar, and two
similar AD lattices, patterned on the $[\bar{2}33]$ arm. The ratio
$d/a$ for each AD cell is $\simeq 1/3$, where $d$ is the AD
diameter. The sample was thinned to $\simeq 200~\mu$m and then glued
to one side of a commercial piezo-electric stack actuator with the
sample's [$01\bar{1}$] crystal direction aligned to the piezo's
poling direction. Voltage biasing the piezo results in in-plane strain which is transmitted to the sample
\cite{ShayeganAPL03,BaburPRB07}. We specify strain ($\epsilon$)
values along the poling direction; in the perpendicular
($[\bar{2}33]$) direction, the strain is approximately
$-0.38\epsilon$ \cite{ShayeganAPL03}. Longitudinal MR was measured,
as a function of perpendicular magnetic field ($B$) at $T = 0.3$~K.
The data reported here were all taken during a single sample
cool-down.
\begin{figure}
%\centering
\includegraphics[scale=1]{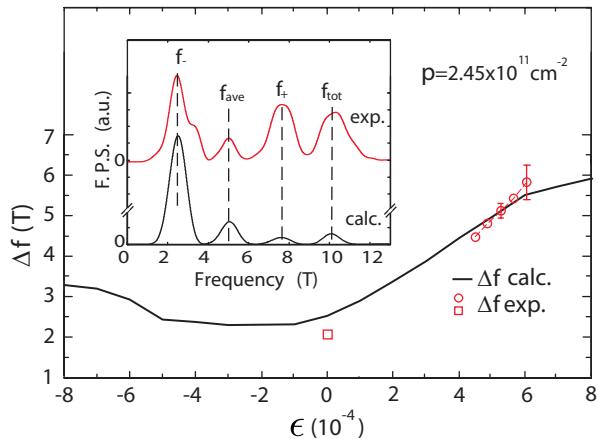}
\caption{\label{fig:fig2}(Color online) The solid curve shows
$\Delta f$($ = f_{+} - f_{-}$), determined from the calculated
magneto-oscillations, as a function of strain. The circles are the
experimentally measured $\Delta f$. The square is the $\Delta f$
measured on a sample from the same wafer not mounted on a piezo.
The inset shows the Fourier power spectra (F.P.S.) of the measured
and calculated magneto-oscillations at $\epsilon=5.3 \times
10^{-4}$.}
\end{figure}

\begin{figure*}
\centering
\includegraphics[scale=0.35]{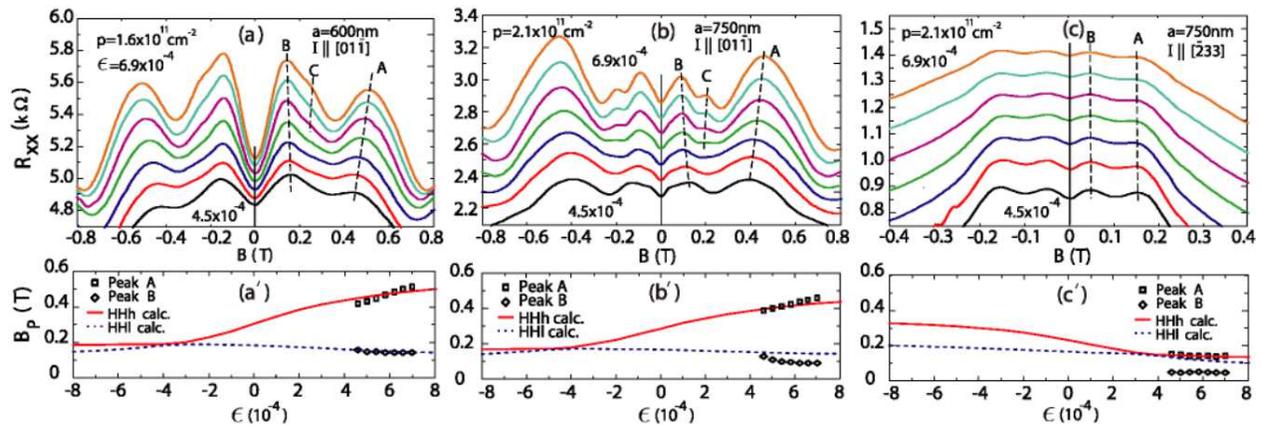}
\caption{\label{fig:fig3}(Color online) (a),(b) and (c):
Low-field magneto-resistance traces for the indicated 2D hole
density ($p$), anti-dot lattice period ($a$), and current
direction. In each figure traces are shown for the indicated
strain values from $6.9 \times 10^{-4}$ (top trace) to $4.5 \times
10^{-4}$ (bottom trace) in steps of $4 \times 10^{-5}$. The
resistance scale is for the lowest trace and the other traces are
offset vertically for clarity. (a$'$),(b$'$) and (c$'$):
Comparison of the expected magneto-resistance peak positions
($B_{P}$) based on Fermi contour calculations and the positions of
the experimentally observed peaks A and B.}
\end{figure*}

%%%%%%%%%%%%%%%%%%%%%%%%%%%%%%%%%%%%%%%%%%%%%%%%%%%%%%%%%%%%%%%%%%
% \section{Experimental Results}

Before presenting the commensurability MR data we first discuss
our measurements of spin-splitting which we use to determine the
amount of residual (built-in) strain in our sample. This residual
strain exists because of the anisotropic contraction of the
piezo-actuator to which the sample is glued, and depends on the
cool-down of the piezo and the sample. We determine this strain by
comparing the measured spin-splitting in our sample to the results
of our self-consistent calculations of 2D hole bands, based on the
$8\times8$ Kane Hamiltonian and augmented by the strain
Hamiltonian of Bir and Pikus \cite{Winkler03,Bir74,TrebinPRB79}.
They take into account the spin-orbit coupling due to both the
structure inversion asymmetry of the GaAs/AlGaAs heterojunction as
well as the bulk inversion asymmetry of the underlying zinc-blende
structure \cite{WinklerPRB93}. These energy band calculations have
successfully explained the spin-orbit induced spin-splitting and
its dependence on density, electric field and strain in 2DHSs
\cite{LuPRL98,PapadakisPHE01,BaburPRB07}. To make a direct
comparison with the experimental data, we calculated the Landau
fan chart as a function of $B$ and determined the
magneto-oscillations of the density of states at the Fermi energy
\cite{Winkler03,WinklerPRL00}. We then calculated the Fourier
power spectrum of these oscillations and obtained the frequencies
$f_{+}$ and $f_{-}$ that correspond to the majority (HHh) and
minority (HHl) spin-subbands. An example of such a spectrum is
shown in the inset of Fig.\ 2 (lower trace). The difference
$\Delta f$ between $f_{+}$ and $f_{-}$ is a measure of
spin-splitting and can be directly compared to the experimentally
determined $\Delta f$ which we obtain from the Fourier transform
of the measured Shubnikov-de Haas (SdH) oscillations; the upper
trace in Fig.\ 2 inset shows such a spectrum.

In Fig.\ 2 we show the calculated $\Delta f$ (solid curve) as a
function of $\epsilon$. We emphasize that the calculations were done
based on the sample structure and there are no fitting parameters.
In this plot we have also included a point (open square) from SdH
oscillations measured on a sample from the same wafer which was not
mounted on a piezo and is therefore free of strain. This data point
agrees well with the calculations. In Fig.\ 2 we also plot $\Delta
f$, determined from the SdH oscillations measured in the unpatterned
regions of the piezo-mounted sample (open circles), where we have
shifted all the measured data points horizontally by $ 5.7 \times
10^{-4}$ in order to match the calculated $\Delta f$. (Note that the
relative strains for these data points are determined from our
calibration of the piezo-actuator, and are known to better than
about $10\%$.) The plot of Fig.\ 2 implies that the sample is under
$\simeq 5.7 \times 10^{-4}$ of residual (tensile) strain. This is
somewhat larger than the typical residual strain values in our
experiments on other samples, but is not unreasonable. In the
remainder of this report, we assume that this is indeed the residual
strain in our sample.

We now describe the results of our low-field MR measurements on
the AD lattice regions of the sample. First we focus on the
results for the $a=600$~nm period region with the current along
[$01\bar{1}$]. The MR traces for this region are shown in Fig.\
3(a) for different values of strain as indicated. We observe two
strong MR peaks, labelled as A and B in the figure. The field
positions of these peaks ($B_{P}$) are plotted in Fig.\ 3(a$'$)
for different values of $\epsilon$. Note that these strain values
are based on Fig.\ 2 data, i.e., a residual strain of $5.7 \times
10^{-4}$ is assumed, but otherwise there are no adjustable
parameters. The data of Figs.\ 3(a) and (a$'$) indicate that peak
A moves to higher fields as the sample is further strained while
peak B shifts to slightly smaller fields. As we discuss below, the
positions of these peaks and their dependence on strain are in
excellent agreement with what we expect from our calculations of
our sample's Fermi contours.

Examples of the calculated HHh and HHl spin-subband Fermi contours
are shown in Fig.~1(a). When the cyclotron orbit diameter in the
direction perpendicular to the current flow equals an integer
multiple of the AD lattice period in that direction, the MR
exhibits a peak \cite{Weiss91, Lorke91, Fleisch92,
TsukagoshiPRB95, LuPRB96, OkiPRB07, ZitzlspergerEPL03}. In other
words, we expect MR peaks at fields $B_{P}=2\hbar k_{F}/eai$ where
$k_{F}$ is the Fermi wavevector along the current direction and
$i=1,2,3, \ldots$ is an integer. In Fig.\ 1(b) we show the
smallest ($i=1$) commensurability orbits for the HHh and HHl holes
that are expected to lead to a MR peak for the current along
[$01\bar{1}$]. In Fig.\ 3(a$'$) we have plotted (solid and dashed
curves) the positions of the MR peaks for these two orbits
expected from the calculations. Obviously, there is an excellent
agreement between the expected peak positions and the
experimentally observed values. This agreement is particularly
remarkable considering that, except for assuming a residual strain
of $5.7 \times 10^{-4}$, there are no adjustable parameters in
comparing the MR peak positions.

We note that in our calculations the anisotropy of the Fermi
contours emerges as follows. In hole systems the spin-splitting is
greatly influenced by the energy separation between the heavy- and
light-hole states \cite{Winkler03}. This splitting depends in turn on the direction of the in-plane
wavevector relative to the orientation of the strain
\cite{TrebinPRB79} which results in the highly anisotropic
spin-dependent deformation of the Fermi contours displayed in
Fig.\ 1(a).

In Fig.\ 4 we show similar data taken on the same AD region
($a=600$~nm) as in Fig.\ 3, but here the MR traces are taken at
different values of hole density while the strain is kept fixed at
$5.7 \times 10^{-4}$. Two strong MR peaks A and B are observed
whose positions move to higher fields with increasing density. As
seen in Fig.\ 4, there is again good quantitative agreement
between the experimental data and the calculated positions of
these MR peaks.

Data for the AD region with period $750$~nm and with current along
[$01\bar{1}$] are shown in Figs.\ 3(b) and (b$'$). The MR traces
are overall qualitatively similar to those for the $600$~nm period
AD region in that they show two strong peaks, labelled A and B,
which move in opposite directions in field with increasing strain.
The positions of these peaks are in good agreement with the
results of the calculations, although peak B is observed at
somewhat lower field values than expected.

In Figs.\ 3(c) and (c$'$) we show the data for the $750$~nm period
region with current along [$\bar{2}33$]. Again we observe two
prominent peaks, which we label A and B, but here the peaks are seen
at very low field values and their positions are essentially
independent of strain. Note that in this case the MR data
probe $k_{F}$ along [$\bar{2}33$] for the HHh and HHl subbands. As seen in Fig.\ 1(a), in the strain range that our
experiments probe ($4.5$ to $6.9 \times 10^{-4}$), $k_{F}$ of the HHh and HHl bands along [$\bar{2}33$] are close
in size and are both smaller than $k_{F}$ along
[$01\bar{1}$]. This is reflected in the $B_{P}$ curves shown in
Fig.\ 3(c$'$) which predict two essentially overlapping MR peaks
that occur at small fields and whose positions are nearly
independent of strain. It is therefore likely that our observed peak
A corresponds to the commensurate orbits of both HHh and HHl bands,
and peak B is the low-field peak commonly observed in AD
lattices with soft potentials \cite{Fleisch92}.

We also took data on another sample from the same wafer and
with similar AD lattices, but not mounted on a piezo-actuator so
that it is strain-free. Our results for this sample are similar to
those reported by Zitzlsperger {\it et al.} \cite{ZitzlspergerEPL03} where MR peaks in an un-strained 2DHS with AD lattices were studied, and
corroborate our conclusions. The low-field MR traces for our
un-strained sample show a peak at a magnetic field whose position
lies between the two $B_{P}$ theoretically expected for the HHh and HHl
contours. Evidently we cannot resolve the two commensurability peaks
for the HHh and HHl bands at zero strain where the two peaks are
theoretically expected to be close to each other. Given the
relatively large width of the MR peaks, this is not
surprising. More revealing is the observation that the MR traces for
the AD lattices patterned on the two Hall bar arms (along
[$\bar{2}33$] and [$01\bar{1}$]) of our strain-free sample are
fairly similar. This implies that the Fermi contours are nearly
isotropic at zero strain, consistent with the calculations (see Fig.\
1(a)). In contrast, the data in Figs.\ 3(b,b') and 3(c,c') are
clearly very different, reflecting the strong anisotropy of the
Fermi contours at large strains.

\begin{figure}
\centering
\includegraphics[scale=1]{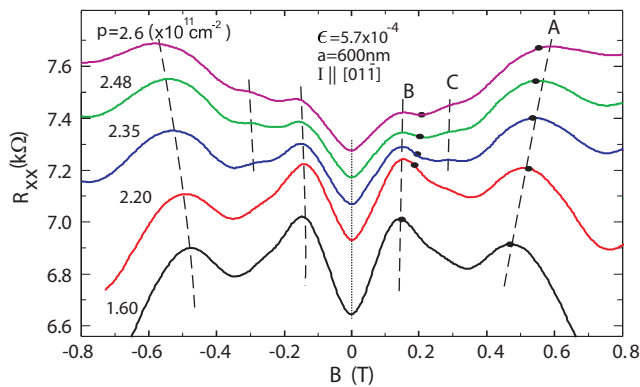}
\caption{\label{fig:fig4}(Color online) Low-field MR traces for
the 600~nm AD region for different densities at $\epsilon = 5.7
\times 10^{-4}$. Traces are offset, except for the bottom one.
Theoretically expected MR peak positions are shown by filled
circles on each trace.}
\end{figure}

While the overall agreement between the experimental data and
calculations shown in Figs.~3 and 4 is good \cite{Footnote2}, there
are some puzzles. First, the positions of the lower MR peak (peak B)
in most of the experimental data are somewhat smaller than expected
from the calculations. Second, the variation (slope) of the peak A
position with strain is slightly faster than the calculations
predict. We do not have a clear explanation for these discrepancies.
They might be arising from the
simplifications of theoretical model that is based on the properties
of the un-patterned 2DHS. It is also possible that the cyclotron
orbits are slightly distorted in the AD lattice regions (compared to
the un-patterned regions) because of the non-ideal (e.g., not
perfectly abrupt) potentials of the ADs \cite{Fleisch92}. Third, for
current along [$01\bar{1}$], some of the experimental MR traces
exhibit a weak peak (labelled C) between peaks A and B (see Figs.\ 3
and 4). Note in Fig.\ 1(a) that in the strain range of our
measurements, $k_{F}$ along [$\bar{2}33$] for the HHh
and HHl bands are quite close to each other. It is tempting to
associate this peak C with a magnetic-breakdown-like phenomenon
where the hole trajectory switches from the HHh to HHl orbit when
these orbits come close to each other in $k$-space. Supporting such
a conjecture is the fact that, in the strain and density range where
we observe peak C, the Fourier power spectra of the SdH oscillations
also show a peak at the frequency $f_\mathrm{ave} = (f_{+} +
f_{-})/2$ (see Fig.\ 2 inset). Such a peak could indicate a
magnetic-breakdown \cite{KeppelerPRL02}.

Despite these puzzles, our results provide strong evidence that we are probing the ballistic
transport in individual spin-subbands. They also show that
we can tune the anisotropy of the Fermi contours, and therefore
the cyclotron orbit trajectories, via the application of strain.
Such tuning may find applications in ballistic spintronic devices,
e.g., it could facilitate the separation of carriers in different
spin subbands in magnetic focusing devices \cite{RokhinsonPRL04}.

We thank the DOE, NSF, and ARO for support and S.\ Misra and B.\
Habib for useful advice.


\begin{thebibliography}{99}
\bibitem{Footnote2}
Data for other densities and current direction, not shown here,
exhibit equally good agreement with calculations.
\end{thebibliography}
\end{document}